\documentclass{emulateapj}
\usepackage{apjfonts,graphics}

\newcommand{\simgt}{\lower.5ex\hbox{$\; \buildrel > \over \sim \;$}}
\newcommand{\simlt}{\lower.5ex\hbox{$\; \buildrel < \over \sim \;$}}

\newcommand{\bi}[1]{\hbox{\boldmath{$#1$}}}

\newcommand{\colskip}{@{\hspace{0.3in}}}

\lefthead{Takada $\&$ White}
\righthead{Tomography of lensing cross power spectra}

\begin{document}

\title{Tomography of lensing cross power spectra}

\author{Masahiro Takada\altaffilmark{1} and Martin White\altaffilmark{2}}
\affil{${}^1$Department of Physics, University of Pennsylvania,
 209 S. 33rd Street, Philadelphia, PA 19104: mtakada@hep.upenn.edu}
\affil{${}^2$Departments of Physics and Astronomy, University of
California, Berkeley, CA 94720: mwhite@astro.berkeley.edu}

\begin{abstract}
By obtaining photometric redshift information, tomography allows us to
cross-correlate galaxy ellipticities in different source redshift bins.
The cross-correlation is non-vanishing because the different bins share
much of the foreground mass distribution from which, over Gpc scales,
the lensing signal is built.
If the redshift bins are thick enough however, the cross-correlations are
insensitive to contamination from the intrinsic alignments of galaxies
since these fall off rapidly on scales larger than a few tens of Mpc.
We forecast how lensing tomography using only the cross-power spectra
can constrain cosmological parameters compared to tomography including
the auto-spectra.  It is shown that the parameter errors are degraded
by only ${\cal O}(10\%)$ for 5 or more source redshift bins.
Thus, the cross-power spectrum tomography can be a simple, model-independent
means of reducing the intrinsic alignment contamination while retaining most
of the constraints on cosmology.
\end{abstract}


\section{Introduction}

As our cosmological knowledge and techniques mature, it becomes
increasingly important to strive for observations and methods which are,
as much as possible, immune to known systematic effects as well as being
statistically powerful.
Nowhere is this more true than in the field of weak gravitational lensing. 
Since its detection by several groups three years ago
(Bacon, Refregier \& Ellis 2000; 
 Kaiser, Wilson \& Luppino 2000;
 Van Waerbeke et al. 2000; 
 Wittman et al. 2000),
weak gravitational lensing by large-scale structure has become a well
established technique, used already to set constraints on the mass
density ($\Omega_{\rm m0}$) and the fluctuation amplitude $(\sigma_8)$
(e.g., see Van Waerbeke \& Mellier 2003 for the current status)
and touted for its potential to constrain cluster scaling relations
(Huterer \& White 2002) and dark energy
(Benabed \& Bernardeau 2001; 
 Huterer 2002; Hu 2002a,b; Heavens 2003; Abazajian \& Dodelson 2003;
 Refregier et al. 2003; Jain \& Taylor 2003; Bernstein \& Jain 2003;
 Takada \& Jain 2003, hereafter TJ03).  

There are many sources of systematic errors which can affect lensing
measurements. 
One source in particular, intrinsic ellipticity alignments of source galaxies
(Heavens, Refregier \& Heymans 2000; Catelan, Kamionkowski \& Blandford 2001;
 Crittenden et al.~2001; Pen, Lee \& Seljak 2001;
 Mackey, White \& Kamionkowski 2002; Jing 2002)
cannot be mitigated by improved instrumentation or reliably predicted by
theory.
While likely not important for the current generation of experiments
(Heymans et al.~2003), 
it could prove to be a limiting uncertainty in more ambitious ongoing and
future surveys such as the
CFHT Legacy Survey, the
Deep Lens Survey, 
DML/LSST,
Pan-STARRS and
SNAP.

In this {\em Letter\/}, we discuss a simple, largely (astrophysical) model
independent, technique to reduce the susceptibility of weak lensing
measurements to intrinsic alignments of galaxies.
This technique assumes only that the intrinsic alignments fall off on
scales above a few tens of Mpc while the lensing signal builds up over
Gpc scales. 
We emphasize the advantages of splitting the galaxy distribution into multiple
redshift bins and considering only the cross-power spectra.
These have no shot noise bias, contain most of the cosmological information
and should be totally insensitive to intrinsic alignments.
While not as powerful as model dependent methods
(Heymans \& Heavens 2003; King \& Schneider 2002, 2003; Heymans et al.~2003)
these latter require modeling of intrinsic alignments, which involves many
uncertain aspects of galaxy formation. 

\section{Tomography of lensing cross-power spectrum}

\subsection{Weak lensing field}

All future surveys will provide photometric redshift information on source
galaxies.  This additional information allows us to subdivide the galaxies
into redshift bins which is crucial if these surveys are to be used to
constrain the evolution of cosmological parameters.  We shall assume
throughout that we have the ability to divide our source population into
redshift bins, commenting on some of the issues in the conclusions.

In the context of cosmological gravitational lensing and assuming the Born
approximation (Blandford et al.~1991; Miralda-Escude 1991; Kaiser 1992,
and see Jain, Seljak \& White 2000 and Vale \& White 2003 for tests),
the lensing convergence field can be expressed as a weighted projection of
the 3D density fluctuation field between observer and source
(also see Bartelmann \& Schneider 2001; Mellier 1999 for reviews):
\begin{equation}
  \kappa(\bi{\theta})=\int_0^{\chi_H} d\chi
  \ W(\chi) \delta[\chi,\chi\bi{\theta}],
\label{eqn:kappai}
\end{equation}
where $\bi{\theta}$ is the angular position on the sky, $\chi$ is the comoving
distance, and $\chi_H$ is the distance to the horizon.  
We shall assume throughout a spatially flat universe. 

The lensing weight function, $W_{(i)}$, for galaxies of subsample $i$ sitting
in a redshift bin $i$ can be written
\begin{eqnarray}
  W_{(i)}(\chi)&=&
  \left\{ \begin{array}{ll} {\displaystyle
  {W_0\over\bar{n}_i}{\chi\over a(\chi)}
  \int_{\chi_{i}}^{\chi_{i+1}}\!\!d\chi_{s}
  p_s(z){dz\over d\chi_s} {\chi_{\rm s}-\chi\over\chi_s}
  }, & \chi\le\chi_{i+1},\\
  0,& \chi>\chi_{i+1}.
  \end{array} \right. 
\label{eqn:weight}
\end{eqnarray}
where $a(\chi)$ is the scale factor, $p_z$ is the redshift distribution of
source galaxies and $W_0\equiv (3/2)\,\Omega_{\rm m0}H_0^2$, with $H_0$ the
Hubble constant and $\Omega_{\rm m0}$ the present day value of the matter
density in units of the critical density. 
Following TJ03, we assume
\begin{equation}
  p_s(z)=\bar{n}_{\rm g}\frac{z^2}{2z_0^3}e^{-z/z_0},
\label{eqn:pz}
\end{equation}
with $z_0=0.5$ and the average number density of galaxies per
steradian, $\bar{n}_{\rm g}$, which peaks at $2z_0=1$ and has median
redshift $z_{\rm med}=1.5$.
The quantity $\bar{n}_i$ is the average number density of galaxies in a
redshift bin $i$, defined to lie between the comoving distances $\chi_i$
and $\chi_{i+1}$:
$  \bar{n}_i=\int_{\chi_i}^{\chi_{i+1}}\!\!d\chi_s~ p_s(z)\frac{dz}{d\chi_s}
  \qquad $, and 
we have assumed sharp subdivisions of the galaxy redshift distribution
for simplicity.

It is worth noting the dependence on the cosmology in these expressions.
The overall shear amplitude is sensitive to $\Omega_{\rm m0}$ and $\sigma_8$,
which explains the status of cosmological constraints derived from current
lensing surveys.
Tomography allows us to recover the redshift evolution of the efficiency,
$W_{(i)}$, and the mass clustering and the evolution of $\chi$ for non
power-law power spectra.
Each of these is sensitive to the equation of state of dark energy,
however the constraints are determined mainly by the dependence of the
lensing efficiency (see e.g.~Figure 2 in Abazajian \& Dodelson 2003).  

\subsection{The power spectrum and its covariance}

Using the flat sky approximation and the Limber approximation
(Limber 1954; Kaiser 1992), the angular power spectrum between the
lensing fields of redshift bins $i$ and $j$ is
\begin{eqnarray}
  C_{(ij)}(l) =\int^{\chi_H}_0\!\!d\chi W_{(i)}\!(\chi) W_{(j)}\!(\chi)
  \chi^{-2}~ P_\delta\!\left(k=\frac{l}{\chi}; \chi\right),
\label{eqn:cltomo}
\end{eqnarray}
where $P_\delta(k)$ is the 3D mass power spectrum. 
The Limber approximation holds well over the angular scales we consider:
$50\le l\le 3000$ (Jain et al. 2000; White \& Hu 2000; Vale \& White 2003). 
For $l\simgt 100$ the major contribution to $C_{(ij)}(l)$ comes from
non-linear clustering (Jain \& Seljak 1997; also see Figure 2 in TJ03) and
we employ the fitting formula of Smith et al.~(2003) for $P_\delta(k)$,
assuming that it can be applied to dark energy cosmologies
(e.g.~White \& Vale 2003).
We note in passing that the issue of accurate power spectra for general
dark energy cosmologies still needs to be addressed carefully
(see Huterer 2002 for related discussion).  

Assuming that the intrinsic ellipticity distribution is uncorrelated between
different galaxies, the observed power spectrum between redshift bins $i$
and $j$ can be written (Kaiser 1992, 1998) 
\begin{eqnarray}
  C^{\rm obs}_{(ij)}(l) =C_{(ij)}(l)+\delta_{ij}
    \frac{\sigma_{\epsilon}^2}{\bar{n}_{i}},
\label{eqn:obscl}
\end{eqnarray}
where $\sigma_\epsilon$ denotes the rms of the intrinsic ellipticities
and the Kronecker delta function accounts for the fact that the cross-power
measurement is not biased by shot noise. 
In addition, the cross-power is totally insensitive to intrinsic ellipticity
alignments, if the bins are much larger than 10 Mpc
(Heymans \& Heavens 2003; King \& Schneider 2002, 2003; Heymans et al.~2003).

Assuming Gaussian statistics, the covariance between the power spectra
$C_{(ij)}(l)$ and $C_{(mn)}(l')$ is 
\begin{equation}
  {\rm Cov} =
  {\delta_{ll'}\over (2l+1)\Delta l f_{\rm sky}}
  \left[C^{\rm obs}_{(im)}C_{(jn)}^{\rm obs}
   +C^{\rm obs}_{(in)}C_{(jm)}^{\rm obs}\right], 
\label{eqn:covps}
\end{equation}
where $f_{\rm sky}$ is the fraction of sky covered, $\Delta l$ is the
bin width and we have suppressed the argument, $l$, of $C^{\rm obs}$.
If we restrict our analysis to angular scales $l\le 3000$ the statistical
properties of the lensing fields are close to Gaussian
(White \& Hu 2000; Cooray \& Hu 2001).
Further the shot-noise will provide the dominant contribution to the
covariance on the small scales where tomography derives most of its
cosmological constraints, strengthening the case for our Gaussian error
assumption.  It is important to note that even if we consider only the
cross-power spectra, with $i\ne j$ and $m\ne n$, the shot noise contributes
to the covariance when $i=m$ and so on.
Hence, the shot noise induces the statistical errors in the cross-power
spectrum measurement, which in turn propagate into cosmological parameters.

\begin{figure*}[t]
\begin{center}
\resizebox{16cm}{!}{\includegraphics{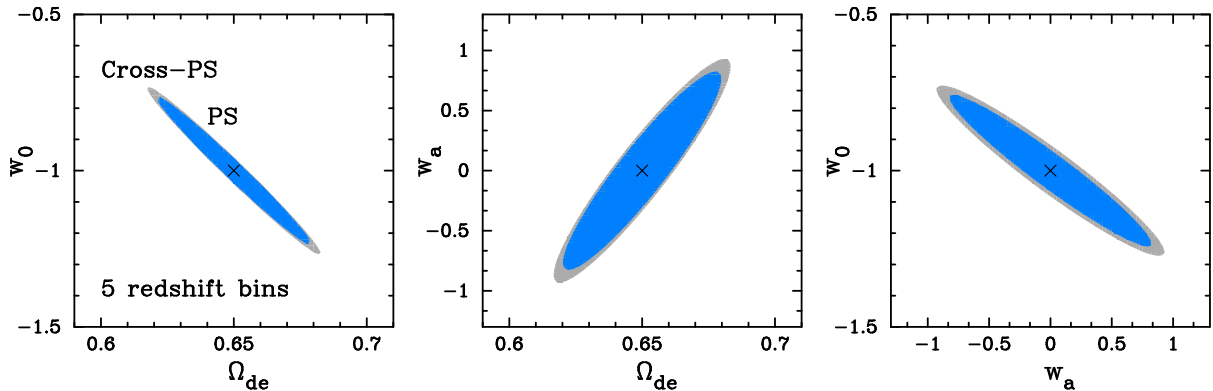}}
\end{center}
\caption{Projected $68\%$ CL constraints in the parameter space of
$\Omega_{\rm de}$, $w_0$, and $w_a$ for 5 bin tomography with and without
the auto-power spectra.}
\label{fig:chins5}
\end{figure*}

\begin{table*}
\begin{center}
Parameter Estimation~ 
($\bar{n}_g=100$ arcmin$^{-2}$, $f_{\rm sky}=0.1$, $\sigma_\epsilon=0.4$)\\
\begin{tabular}{l\colskip cc\colskip cc\colskip cc\colskip cc}\hline\hline
&\multicolumn{2}{c}{\hspace{-2em}$\sigma(\Omega_{\rm de})$} 
&\multicolumn{2}{c}{\hspace{-2em}$\sigma(w_0)$} 
&\multicolumn{2}{c}{\hspace{-2em}$\sigma(w_a)$}
&\multicolumn{2}{c}{\hspace{-2em}$\sigma(\sigma_8)$}\\
redshift bin \# & PS & Cross-PS & PS & Cross-PS & PS & Cross-PS
 & PS & Cross-PS \\ \hline
$n_s=3$& 0.024& 0.057
&0.20&0.51
&0.65&1.5
&0.025& 0.055\\
$n_s=4$&0.021&0.026
&0.17&0.21
&0.58&0.71
&0.022&0.027\\
$n_s=5$&0.020&0.022
&0.16&0.18
&0.55&0.62
&0.020&0.023\\
$n_s=5$ + Bisp&8.2$\times 10^{-3}$& 8.7$\times 10^{-3}$
&0.078&0.084
&0.27&0.29
&$8.3\times 10^{-3}$&8.8$\times 10^{-3}$\\
$n_s=6$&0.019&0.021
&0.15&0.17
&0.52&0.58
&0.020&0.022\\
$n_s=10$&0.018&0.019
&0.15&0.15
&0.49&0.52
&0.019&0.020\\ \hline
\end{tabular}
\end{center}
\caption{Summary of parameter constraints from tomography of only using
 the cross-power spectra (Cross-PS) or tomography including the
 auto-spectra (PS).
 The errors include marginalization over the other parameters.
}  \label{tab:const}
\end{table*}
Table \ref{tab:const} compares parameter forecasts from tomography with
and without the auto-spectra.
We consider angular scales $50\le l\le 3000$, and assume $f_{\rm sky}=0.1$,
$\bar{n}_g=100 $ arcmin$^{-2}$ and $\sigma_\epsilon=0.4$.  
Note that all the errors scale with the sky coverage as $f_{\rm sky}^{-1/2}$.
Given tomography with $n_s$ redshift bins, each bin is chosen so as to have
an equal number density of galaxies for the redshift distribution
(\ref{eqn:pz}); e.g., $0\le z_1\le 1, 1\le z_2\le 1.7, 1.7\le z_3$
for three redshift bins. 
For all the cases we have considered the redshift bins are sufficiently
thick, $\Delta z\simgt 0.2$, that contamination from intrinsic alignments
will be significantly reduced. 

\subsection{Fisher matrix analysis and the fiducial model}
\label{fisher}

Following TJ03, we use the Fisher matrix formalism to examine how lensing
tomography can constrain cosmological parameters.
Assuming the likelihood function for the lensing power spectrum to be
Gaussian, the Fisher matrix can be expressed as
\begin{eqnarray}
F_{\alpha \beta}=\sum_{l=l_{\rm min}}^{l_{\rm max}} \ 
\sum_{(i,j),(m,n)}
\frac{\partial C_{(ij)}(l)}{\partial p_\alpha}
\left[{\rm Cov}\right]^{-1}
\frac{\partial C_{(mn)}(l)}{\partial p_\beta},
\label{eqn:psfis}
\end{eqnarray}
where the inverse covariance matrix is denoted by $[{\rm Cov}]^{-1}$
with ${\rm Cov}$ given by equation (\ref{eqn:covps}) and the $p_\alpha$
denote the cosmological parameters.
To use only the cross-power spectra we simply omit $i=j$ and $m=n$ in
the summation of equation (\ref{eqn:psfis}).
For $n_s$ redshift bins, we can use $n_s(n_s-1)/2$ cross-spectra.
Hence, to extract redshift evolution of the lensing observables requires
at least 3 redshift bins.

The parameter forecasts derived are sensitive to the parameter space
used and to whether constraint on a given parameter is obtained by
marginalizing over other parameter uncertainties.
We use seven parameters to which the lensing observables are sensitive
within the cold dark matter (CDM) model: $\Omega_{\rm de}$, $w_0$, $w_a$,
$\sigma_8$, $n_s$ $\Omega_{\rm b}h^2$ and $h$, where $\Omega_{\rm de}$
and $\Omega_{\rm b}$ are the density parameters of dark energy and baryons,
$n_s$ is the spectral index of primordial scalar perturbations, $h$ is
the Hubble parameter, and $\sigma_8$ is the rms mass fluctuation in a
sphere of radius $8h^{-1}$Mpc. 
We use a simple parameterization of the equation of state of dark
energy: $w(a)=w_0+w_a(1-a)$ (Turner \& White 1997; Linder 2003) and assume
a spatially flat universe.
For the input linear mass power spectrum, we employ the BBKS transfer
function (Bardeen et al.~1986) with the shape parameter given by
Sugiyama (1995).

For the fiducial model we take
$\Lambda$CDM model with $\Omega_{\rm de}=0.65$, $w_0=-1$, $w_a=0$,
$\sigma_8=0.9$, $n_s=1$, $\Omega_{\rm b}=0.05$ and $h=0.72$,
which is consistent with the recent WMAP results (Spergel et al.~2003).  
We assume the priors $\sigma(\ln \Omega_{\rm b}h^2)=0.010$,
$\sigma(n_s)=0.008$ and $\sigma(h)=0.13$, expected from the Planck mission
(see Table 2 in Eisenstein et al.~1999).  
Assuming Gaussian priors, we add the diagonal component
$(F_{\rm prior})_{\alpha\beta}=\delta_{\alpha\beta} \sigma(p_\alpha)^{-2}$
to the lensing Fisher matrix. 

\section{Results}

The table shows that, for three redshift bins, cross-power spectrum tomography
more than doubles the parameter errors from tomography including the
auto-spectra, because we can use only 3 of the 6 spectra.
However, adding even one redshift bin drastically improves the constraints.
As a result, cross-power spectrum tomography for $n_s\ge 5$ recovers most
of cosmological information contained in the auto- and cross-power spectra:
degradation in the $1$-$\sigma$ errors are as small as $\simlt 15\%$.
%
Figure \ref{fig:chins5} shows the constraint ellipses in the parameter space
of $\Omega_{\rm de}$, $w_0$ and $w_a$ for five redshift bins. 
Note that the ellipses correspond to $68\%$ confidence level
($\Delta \chi^2=2.3$), including marginalization over the other parameters.
In this case we can use 10 cross-power spectra for the tomography. 
It is clear that degeneracy directions in the parameter space are almost
identical between the two tomography methods, and the ellipse areas differ 
by only $\sim 30\%$.

TJ03 showed that using both power spectrum and bispectrum tomography
provides improvements in parameter constraints of a factor of 3 over
power spectrum tomography alone.  This is because significant additional
information is contained in the non-Gaussian nature of the large scale
structure inducing the weak lensing.
While $n_s^3$ bispectra can be constructed from $n_s$ redshift bins for
tomography, only the $n_s$ auto-bispectra are contaminated from intrinsic
alignments, allowing us to use the other $n_s(n_s^2-1)$ bispectra for
cross-bispectrum tomography.
The column labeled with ``+ Bisp'' in Table \ref{tab:const} compares
the results when we use all the power spectra and bispectra or use only
the cross-spectra for five redshift bins.
As expected, cross-bispectrum tomography loses little cosmological information
and significantly improves parameter errors derived from the cross-power
spectrum tomography.

\section{Conclusion and Discussion}

We have proposed lensing tomography that only uses the cross-spectra
constructed from source galaxies in different redshift bins.
Cross-spectrum tomography provides a simple, robust and model independent
means of reducing systematics from intrinsic alignments which retains
most of the cosmological information.
For more than 4 redshift bins, cross-spectrum tomography yields errors on
cosmological parameters only $\simlt 15\%$ larger than errors including
the auto-spectra (see Table \ref{tab:const}). 
The situation is further improved by combining with cross-bispectrum
tomography.

As future experiments concentrate on subdividing the source galaxies into
multiple redshift bins to better reconstruct the lensing effect as a
function of distance, it becomes more important to guard against contamination
by intrinsic alignments.
Intrinsic alignments are difficult to predict, with calculations differing
by an order of magnitude at present.
However, simply neglecting intrinsic alignments could bias estimates of
cosmological parameters from future lensing surveys (Heymans et al.~2003). 
Our suggestion is intermediate between detailed modeling of the effect
and simple neglect.
We believe it is worth exploring a 3D mass reconstruction technique using
only the lensing effects on galaxies in different redshift bins 
even if this
is only used as a `sanity check' during the analysis.

Finally we remark on some issues regarding photometric redshift determinations.
In order to implement our procedure we require more redshift bins than is
usual for tomographic techniques (which frequently saturate at 2 or 3 bins).
This requires more accurate photo-$z$ determinations, but still well within
the range of possible accuracies.  For example, {\sl SNAP\/} is designed to
achieve accurate photo-$z$ measurement with a random error $\sigma_{z}=0.03$
(Massey et al.~2003).
This would allow us to perform tomography with up to 10 redshift bins
in the range $0<z<3$.
Of course it is still crucial to calibrate the photo-$z$ estimates to the
percent level, since constraints on the cosmological parameters come from
measuring the lensing efficiency to the percent level.
This will require a spectroscopic survey to calibrate the photo-$z$
distribution, as discussed in Bernstein \& Jain (2003).
Such a measurement should also mitigate against another important
systematic arising from the tails of the photo-$z$ distribution:
the misidentification as source galaxies of a fraction of galaxies which
actually lie near the lensing plane leading to a misestimation of the
lensing efficiency.
Both effects can be controlled introducing gaps, larger than the photo-$z$
error, in the galaxy distribution when computing the cross power spectra
between neighboring bins. 
We have found that even gaps with $\Delta z=0.2$, a typical photo-$z$ error
from a five color survey, enlarge the errors from cross spectrum tomography
with 5 redshift bins by less than 5\%, as expected from our earlier
discussion.
While further study (including realistic photo-$z$ errors) is needed,
this issue is not significantly different for cross-spectra and auto-spectra. 

\acknowledgments
We thank B.~Jain, G.~Bernstein, M.~Jarvis and A.~Heavens for valuable
discussions and comments.
MT would like to thank J.~Cohn, G.~Smoot and the Berkeley Cosmology Group
for their warm hospitality while this work was initiated.
This work was supported by NASA and the NSF.


\begin{thebibliography}{}

\bibitem[]{Kev} Abazajian, K.~ N., Dodelson, S., 2003, Phys.~Rev.~Lett.,
91, 041301

\bibitem[]{Bacon}
Bacon, D.~J., Refregier, A.~R., Ellis, R.~S., 2000, \mnras, 
318, 625

\bibitem[Bardeen]{BBKS}
Bardeen, J.~M., Bond, J.~R., Kaiser, N., 
Szalay, A.~S., 1986, \apj, 304, 15

\bibitem[Bartelmann \& Schneider 2001]{BS01}
Bartelmann, M., Schneider, P., 2001, Phys.~Rep. 340, 291

\bibitem[]{Bernardeau}
Benabed, K., Bernardeau, F., 2001, \prd, 64, 3501

\bibitem[]{Gary}
Bernstein, G., Jain, B., 2003, \apj~ in press, astro-ph/0309332

\bibitem[]{Blandford}
Blandford, R.~D., Saust, A.~B., Brainerd, T.~G., Villumsen, J.~V., 
1991, \mnras, 251, 600

\bibitem[]{Boughn}
Boughn, S.~P., Crittenden, R.~G., 2003, astro-ph/0305001

\bibitem[]{Catelan}
Catelan, P., Kamionkowski, M., Blandford, R.~D., 2001, \mnras, 320, L7

\bibitem[]{CooHu}
Cooray A.R., Hu W., 2001, \apj, 554, 56

\bibitem[]{Crittenden}
Crittenden, R.~G., Natarajan, P., Pen, U.~L., Theuns, T., 2001
\apj, 559, 552

\bibitem[]{Eisenstein}
Eisenstein, D.~J., Hu, W., Tegmark, M., 1999, \apj, 518, 2

\bibitem[]{Heavens}
Heavens, A., 2003, astro-ph/0304151

\bibitem[]{HRH}
Heavens, A., Refregier, A., Heymans, C., 2000, 

\bibitem[]{Heymans}
Heymans, C., Heavens, A., 2003, \mnras, 339, 711

\bibitem[]{Heymans03}
 Heymans,  C., 
 Brown, M., Heavens, A., Meisenheimer, K., Taylor, A., Wolf, C.,
 2003, astro-ph/0310174



\bibitem[]{Hu02a}
Hu, W., 2002a, \prd, 65,   023003



\bibitem[]{Huterer}
Huterer, D., \prd, 2002, 65, 3001

\bibitem[]{HutWhi}
Huterer, D., White, M., 2002, \apj, 578, L95


\bibitem[]{Jain03}
Jain, B., Taylor, A., 2003, \prl, 91, 141302

\bibitem[]{Jain97}
Jain, B., Seljak, U., 1997, \apj, 484, 560

\bibitem[Jain, Seljak \& White 2000]{JSW00}
Jain, B., Seljak, U., White, S.~D.~M., 2000, ApJ, 530, 547

\bibitem[]{Jing}
Jing, Y.~P., 2002, \mnras, 335, L89

\bibitem[]{Kaiser92}
Kaiser, N., 1992, \apj, 388, 272

\bibitem[]{Kaiser98}
Kaiser, N., 1998, \apj, 498, 26

\bibitem[]{Kaiser00}
Kaiser, N., Wilson, G., Luppino, G., 2000, astro-ph/0003338

\bibitem[]{King2002}
King, L.~J., Schneider, P., 2002, A\&A, 396, 411

\bibitem[]{King2003}
King, L.~J., Schneider, P., 2003, A\&A, 398, 23

\bibitem[Limber 1954]{Limber54}
  Limber, D., 1954, \apj, 119, 655

\bibitem[]{Linder}
Linder, E., 2003, \prl, 90, 091301


\bibitem[]{Mackey}
Mackey, J., White, M., Kamionkowski, M., 2002, 
\mnras, 332, 788

\bibitem[]{Massey}
 Massey, R., et al., 2003, astro-ph/0304418

\bibitem[Mellier 1999]{Mellier99}
  Mellier, Y., 1999, ARAA, 37, 127

\bibitem[Pen et al. 2001]{Pen}
  Pen, U.~L., Lee, J., Seljak, U., 2001, \apj, 543, L107 

\bibitem[]{Ref}
Refregier, A. et al., 
 2003, astro-ph/0304419

\bibitem[]{Smith03}
Smith, R.~E. et al., 2003, MNRAS, 341, 1311  (Smith03)

\bibitem[]{Spergel03}
Spergel, D.~N., et al., 2003, ApJ~Suppl., 148,  175

\bibitem[]{Sugiyama}
Sugiyama, N., 1995, \apj~ Suppl., 100, 281

\bibitem[]{TJ03c}
Takada, M., Jain, B., 2003, astro-ph/0310125 (TJ03)


\bibitem[]{TW}
Turner, M.~ S., White, M., 1997, \prd, 56, R4439

\bibitem[]{Chris}
Vale, C., White, M., 2003, \apj, 592, 699

\bibitem[]{Lud}
Van Waerbeke, L., et al., 2000, A\&A, 358, 30

\bibitem[]{Lud03}
Van Waerbeke, L., Mellier, Y., 2003, astro-ph/0305089

\bibitem[]{WH}
White, M., Hu, W., 2000, \apj, 537, 1

\bibitem[]{WhiVal}
White, M., Vale, C., 2003, submitted to Phys. Rev. D. [astro-ph/0312133]

\bibitem[]{Wittman}
Wittman, D.~M., Tyson, J.~A., Kirkman, D., Dell'Antonio, I., 
Bernstein, G., 2000, \nat, 405,  143
\end{thebibliography}
\end{document}